\RequirePackage{amsthm}

\documentclass[pdflatex,sn-mathphys,Numbered,iicol]{sn-jnl}

\usepackage{graphicx}%
\usepackage{multirow}%
\usepackage{amsmath,amssymb,amsfonts}%
\usepackage{amsthm}%
\usepackage{mathrsfs}%
\usepackage[title]{appendix}%
\usepackage{textcomp}%
\usepackage{manyfoot}%
\usepackage{booktabs}%
\usepackage{algorithmicx}%
\usepackage{algpseudocode}%
\usepackage{listings}%

\usepackage{subcaption}
\usepackage{caption}
\usepackage{natbib}
\usepackage{soul}
\usepackage{color}
\usepackage[dvipsnames,table]{xcolor}
\usepackage{pifont}
\usepackage{graphicx}
\usepackage{svg}
\usepackage[inline]{enumitem}
\usepackage{array}
\usepackage{longtable}
\usepackage{tabularx}
\usepackage{graphicx}
\usepackage{makecell}
\usepackage{lscape}
\usepackage{threeparttable}
\usepackage{xurl}
\usepackage{wasysym}
\usepackage{tikz}
\newcommand*\circled[1]{\tikz[baseline=(char.base)]{
            \node[shape=circle,draw,inner sep=1pt] (char) {#1};}}

\usepackage{algorithm}
\usepackage{algpseudocode}

\usepackage[
  separate-uncertainty = true,
  multi-part-units = repeat
]{siunitx}

\usepackage{tablefootnote}
\usepackage{footnote}
\makesavenoteenv{tabular}
\makesavenoteenv{table}
\makesavenoteenv{table*}

\setstcolor{blue}

\newtoggle{finalPaper}
\toggletrue{finalPaper} 

\iftoggle{finalPaper} {
    \newcommand{\addtxt}[1]{#1}
    \newcommand{\addtable}[1]{#1}
    \newcommand{\change}[2]{#2}
    \newcommand{\rmvtxt}[1]{}
} {
    \newcommand{\addtxt}[1]{\textcolor{blue}{#1}}
    \newcommand{\addtable}[1]{\color{blue}{#1}}
    \newcommand{\change}[2]{\st{#1}\textcolor{blue}{#2}}
    \newcommand{\rmvtxt}[1]{\st{#1}}
}

\definecolor{ao}{rgb}{0.0, 0.5, 0.0}
\definecolor{amber}{rgb}{1.0, 0.49, 0.0}
\newcommand{\yestick}{{\color{ao}\ding{51}}}
\newcommand{\notick}{{\color{red}\ding{55}}}

\newcommand\sbullet[1][.5]{\mathbin{\vcenter{\hbox{\scalebox{#1}{$\bullet$}}}}}

\definecolor{gray45}{gray}{.45}
\definecolor{gray75}{gray}{.75}
\definecolor{orange-fig}{HTML}{C55A11}

\theoremstyle{thmstyleone}%
\theoremstyle{thmstyletwo}%

\theoremstyle{thmstylethree}%

\begin{document}


\title[Mitigating Communications Threats in Decentralized Federated Learning through Moving Target Defense]{Mitigating Communications Threats in Decentralized Federated Learning through Moving Target Defense}


\author*[1]{\fnm{Enrique Tomás} \sur{Martínez Beltrán}}\email{enriquetomas@um.es}
\author[1]{\fnm{Pedro Miguel} \sur{Sánchez Sánchez}}\email{pedromiguel.sanchez@um.es}
\author[1]{\fnm{Sergio} \sur{López Bernal}}\email{slopez@um.es}
\author[2]{\fnm{Gérôme} \sur{Bovet}}\email{gerome.bovet@armasuisse.ch}
\author[1]{\fnm{Manuel} \sur{Gil Pérez}}\email{mgilperez@um.es}
\author[1]{\fnm{Gregorio} \sur{Martínez Pérez}}\email{gregorio@um.es}
\author[3]{\fnm{Alberto} \sur{Huertas Celdrán}}\email{huertas@ifi.uzh.ch}

\affil[1]{\orgdiv{Department of Information and Communications Engineering}, \orgname{University of Murcia}, \orgaddress{\postcode{30100}, \country{Spain}}}
\affil[2]{\orgdiv{Cyber-Defence Campus}, \orgname{Armasuisse Science and Technology}, \orgaddress{\postcode{3602}, \city{Thun}, \country{Switzerland}}}
\affil[3]{\orgdiv{Communication Systems Group, Department of Informatics (IFI)}, \orgname{University of Zurich}, \orgaddress{\postcode{8050}, \city{Zürich}, \country{Switzerland}}}

\abstract{The rise of Decentralized Federated Learning (DFL) has enabled the training of machine learning models across federated participants, fostering decentralized model aggregation and reducing dependence on a server. However, this approach introduces unique communication security challenges that have yet to be thoroughly addressed in the literature. These challenges primarily originate from the decentralized nature of the aggregation process, the varied roles and responsibilities of the participants, and the absence of a central authority to oversee and mitigate threats. Addressing these challenges, this paper first delineates a comprehensive threat model\change{, highlighting the potential risks of}{ focused on} DFL communications. In response to these identified risks, this work introduces a security module to counter communication-based attacks for DFL platforms. The module combines security techniques such as symmetric and asymmetric encryption with Moving Target Defense (MTD) techniques, including random neighbor selection and IP/port switching. The security module is implemented in a DFL platform, Fedstellar, allowing the deployment and monitoring of the federation. A DFL scenario \change{has been deployed, involving eight physical devices implementing}{with physical and virtual deployments have been executed, encompassing} three security configurations: (i) a baseline without security, (ii) an encrypted configuration, and (iii) a configuration integrating both encryption and MTD techniques. The effectiveness of the security module is validated through experiments with the MNIST dataset and eclipse attacks. \change{The results indicated an average F1 score of 95\%, with moderate increases in CPU usage (up to 63.2\% $\pm$3.5\%) and network traffic (230 MB $\pm$15 MB) under the most secure configuration, mitigating the risks posed by eavesdropping or eclipse attacks.}{The results showed an average F1 score of 95\%, with the most secure configuration resulting in CPU usage peaking at 68\% ($\pm$9\%) in virtual deployments and network traffic reaching 480.8 MB ($\pm$18 MB), effectively mitigating risks associated with eavesdropping or eclipse attacks.}}


\maketitle


\section{Introduction}\label{sec:introduction}

The rise of the Internet of Things (IoT) has significantly reshaped the digital landscape, defining an era marked by unprecedentedly interconnected devices. IoT devices produce vast volumes of data every second, spanning various sectors, from healthcare and manufacturing to transportation and home automation. Traditionally, Machine Learning (ML) techniques have been employed to derive meaningful insights from these large datasets. However, these techniques often involve the centralized aggregation of data, a process that raises serious concerns about data privacy, data sovereignty, and overhead \cite{idc:iot_number:2022}.

A novel ML approach, known as Federated Learning (FL), has emerged in response to these challenges. FL can train models locally on multiple edge devices, each holding local data samples. This eliminates the need to share raw data, thereby preserving data privacy. \change{Furthermore}{Advancing this concept}, Decentralized Federated Learning (DFL) represents a paradigm shift within FL \cite{MartinezBeltran:DFL_survey:2023}. \change{DFL enhances decentralization by facilitating model aggregation at multiple nodes, drastically reducing dependence on a single central server and enabling direct, pairwise model sharing among the network nodes.}{DFL strengthens decentralization by enabling the aggregation of models across multiple nodes, thereby substantially reducing reliance on a centralized server. This advancement not only preserves privacy but also enhances system scalability, robustness, and efficiency, making it particularly suitable for distributed IoT applications. DFL integrates several key processes: (1) participants train models on their edge devices using local data, preserving data privacy; (2) nodes then directly exchange the parameters of their models in pairs, which favors a decentralized network structure; and (3) each node integrates these shared parameters into their local models, resulting in an aggregated and refined model that benefits from the diverse data insights from across the network.} This innovative approach addresses single points of failures, trust dependencies, and server node bottlenecks inherent in traditional FL. DFL also eliminates the need for a central server by broadening the model aggregation to multiple nodes. Additionally, DFL employs asynchronous communications, a departure from traditional FL. This feature enables individual nodes to communicate their updates independently of others, contributing to system resilience and ensuring the continued learning process even if some nodes encounter delays or disconnections \cite{Shi:personalization_dfl:2023}. The application of DFL to wireless networks has been motivated by the resilience offered by its asynchronous communication, which is crucial in environments with intermittent and unpredictable connectivity \cite{Salama:dfl_wireless:2023}. Specifically, such traits make DFL highly applicable for Unmanned Aerial Vehicle (UAV) networks, where constant and reliable communication is often challenged by diverse factors such as terrain and weather conditions, hence enhancing their cooperative missions \cite{Xiao:military_fullydl:2021}.

Despite the substantial benefits of DFL, it also introduces new challenges. This approach poses different types of sensitive information necessary for the federation, such as the network topology, the roles of the participants, and communication patterns that can be exploited. Besides, in DFL environments where all participants are connected, the absence of a central authority to manage potential threats raises significant security and privacy concerns \cite{PeralesGomez:temporalfed:2023}. With each participant sharing equal threat exposure, adversarial and communication-based attacks become significant concerns. Adversarial attacks can misguide the learning process by manipulating training data or leveraging the shared model updates to infer sensitive information about the other participants. At the same time, communication-based threats can disrupt the model aggregation process or lead to security breaches and privacy infringements \cite{Mothukuri:survey_topology_architecture_privacy_decentralized:2021}. Addressing these challenges could benefit from adopting a dynamic approach like Moving Target Defense (MTD) \cite{Etxezarreta:MTD_wireless:2023}. MTD is a security concept that continuously alters attack surfaces to confuse and mislead adversaries, making it difficult for them to launch successful attacks. \change{Although not yet extensively validated in the context of DFL, the potential integration of MTD with other security techniques could improve the overall security posture of the system against evolving threats.}{The potential integration of MTD with encryption in DFL offers a novel approach to enhancing security, particularly in the face of unique challenges in decentralized architectures. This strategy is particularly relevant in DFL, where the decentralized nature of data exchange and interaction presents distinct challenges not adequately addressed by traditional security methods. Combining dynamic MTD techniques with strong encryption proposes an advanced defense against vulnerabilities and threats unique to these systems. Moreover, the literature has not extensively addressed specific attacks within DFL environments, highlighting the need for this innovative integration. Such an approach underscores the need for innovative solutions tailored to environments with distributed architectures.} In recognition of the risks in DFL, and with a special emphasis on communication-based attacks that leverage the inherent decentralization of DFL, this paper presents the following contributions:

\begin{itemize}

    \item Create a threat model, identifying and understanding the sensitive information vulnerable to threats affecting the communications in DFL, such as eavesdropping, Man in the Middle (MitM), and eclipse attacks.
    
    \item Develop an advanced security module for DFL platforms providing secure data exchanges through encryption and dynamic proactive defense using MTD. This module mitigates the threats identified in the comprehensive threat model \addtxt{of DFL}, ensuring efficient system operation despite the integrated security measures.

    \item Implement and deploy the security module within a real-world DFL framework, Fedstellar, integrating it into the frontend, controller, and core components of the platform to enhance the overall security of the DFL approach. Furthermore, this work \change{establishes a realistic DFL environment using the Fedstellar platform, consisting of eight interconnected heterogeneous physical devices.}{implements a dual DFL environment using the Fedstellar platform. The initial deployment comprises a physical network of eight heterogeneous devices. Additionally, a virtual deployment with 50 participants facilitates a comprehensive and scalable evaluation of DFL performance.} Three security configurations are assessed in \change{this setup}{both setups}: a baseline with no security, a configuration with encryption, and a configuration integrating both encryption and MTD techniques.
    
    \item Conduct an in-depth experimental evaluation of the proposed security module using a real-world topology with diverse connections and participants, leveraging the widely used MNIST dataset and \addtxt{a custom implementation of an} eclipse attack. \change{The evaluation shows an average F1 score of 95\%, reaching 97\% without security measures. Secure configurations, specifically those employing encryption and MTD, induce a slight increase in CPU usage (to 63.2\% $\pm$3.5\%) and network traffic (to 230 MB $\pm$15 MB). Meanwhile, there is a minor RAM rise, peaking at 33.9\% ($\pm$1.5\%) under the encryption and MTD configuration.}{The evaluation across both physical and virtual deployments reveals an average F1 score of 95\%, which ascends to 98.9\% in the absence of security measures. Implementing secure configurations, particularly those utilizing encryption and MTD, leads to an increase in CPU usage, reaching up to 68\% ($\pm$9\%) in virtual environments. In addition, the network traffic peaks at 480.8 MB ($\pm$18 MB), while the RAM usage also experiences a moderate rise, with a maximum of 35.9\% ($\pm$1.5\%) noted in the physical deployment under encryption and MTD settings.}
\end{itemize}

The remainder of this paper is organized as follows: Section~\ref{sec:relatedwork} provides an in-depth overview of the literature on DFL and its associated security challenges. Section~\ref{sec:threatmodel} introduces the proposed threat model, highlighting the unique security issues that DFL environments face. Section~\ref{sec:module} presents a detailed description of the proposed security module, elucidating its key components and their functionality. Section~\ref{sec:scenario} outlines the experimental setup and evaluation methodology, paving the way for a rigorous assessment of the effectiveness of the security module. Section~\ref{sec:results} presents a comprehensive discussion of the results, and Section~\ref{sec:conclusion} concludes the paper with a summary of the key findings and an exploration of potential avenues for future research.

\section{Related Work}\label{sec:relatedwork}

This section gives the insights required to understand the concepts used in the following sections and reviews the main works in the literature associated with the present one.

\subsection{Privacy and security in DFL}

The promise of DFL as a tool for collaborative learning in heterogeneous and geographically distributed settings continues to drive robust research into its inherent security implications. A comprehensive understanding of its potential threats and appropriate countermeasures enhances cooperative learning practices. Several ground-breaking research efforts have focused on integrating trust within a DFL context. In this regard, Gholami et al. \cite{Gholami:trusted:2022} proposed an approach that integrates trust as a metric within a DFL context. Their method used a comprehensive mathematical framework to quantify and aggregate the trustworthiness of individual agents. In parallel, Mothukuri et al. \cite{Mothukuri:trusted_blockchain:2022} addressed anomaly detection in Internet of Things (IoT) networks by leveraging the distributed nature of FL. They proposed a FL methodology that optimized anomaly detection by aggregating updates from diverse sources. Their approach hinged on using gated recurrent units (GRUs) in federated training rounds to maximize the accuracy of the overall ML model. Complementing these advancements, Li \cite{Li:trust_dfl:2023} took an innovative leap by proposing a Trustiness-based Hierarchical Decentralized FL (TH-DFL) framework. It employs a Security Robust Aggregation (SRA) rule to ensure privacy and robustness even in the face of malicious nodes. The TH-DFL framework strikes an optimal balance between privacy and robustness, especially as the group size fluctuates, and exhibits superior resilience against varying forms of attacks.

Security concerns related to jamming attacks have also been extensively studied, especially in wireless networks implementing DFL. Shi et al. [3] shed light on the susceptibility of DFL to these attacks, proposing crucial countermeasures. Their algorithms identify and target pivotal network links for attack prevention and optimal placement of jammers to disrupt the federation process. Their findings point to the urgency for sophisticated defense mechanisms in DFL architectures. Further contributing to the body of knowledge on security threats in DFL, Chen et al. \cite{Chen:attacks_detection_historical_gradient:2023} proposed a method called Decentralized FL Historical Gradient (DFedHG). DFedHG utilizes historical gradients to differentiate between regular, untrusted, and malicious users in a DFL environment. This unique solution strengthens the defense against potential threats in DFL systems, accentuating the necessity for sturdy security frameworks. 

Securing wireless networks while implementing DFL is a topic of intensive research. Wang \cite{Wang:dfl_secure_wireless:2023} introduced a method to ensure the security and efficiency of FL in Wireless Computing Power Networks (WCPNs). Their research presents a secure and decentralized FL solution based on blockchain for WCPN, which allows nodes to freely participate or leave the WCPN federated training without authorization and security threats. This approach uses a blockchain with a proof-of-accuracy (PoAcc) consensus scheme and an evolutionary game-based incentive scheme to ensure the consistency and security of FL in WCPN. On the other hand, Salama \cite{Salama:dfl_wireless:2023} proposed a method for Decentralized FL over Slotted ALOHA Wireless Mesh Networking. The approach offers an efficient solution for ML model training without a central server, reducing communication costs and increasing convergence speed. This paper demonstrates how network topologies can impact the performance of ML models, and their results indicate significant promise for DFL in Internet of Things (IoT) systems.

\subsection{Security-based DFL solutions}

Innovative approaches toward enhancing data protection and secure communication within DFL environments have also seen considerable development. For instance, the FusionFedBlock solution, proposed by Singh et al. \cite{Singh:bl_5g_dfl:2023}, merges the strengths of blockchain and DFL to ensure privacy in Industry 5.0. A distributed hash table (DHT) guarantees secure decentralized storage at the cloud layer, while blockchain miners facilitate data verification. FL-SEC, introduced by Qu et al. \cite{Qu:privacy_framework_iot:2022}, stands as a breakthrough framework that addresses potential information leakage due to inference attacks, threats of poisoning attacks via falsified data, and high consumption of communication resources. This model uses a custom incentive mechanism and an enhanced sign gradient descent method to protect the privacy of model parameters and significantly reduce communication resource consumption. Contributing further to privacy preservation and trustworthiness in DFL, Wang \cite{Wang:dfl_privacy_trust:2023} proposed PTDFL, an efficient and novel DFL scheme. This scheme integrates a gradient encryption algorithm to protect data privacy, employs concise proof for the correctness of the gradients, and uses a local aggregation strategy to ensure that the aggregated result is trustworthy. The unique feature of PTDFL is its support for data owners joining in and dropping out during the entire DFL task.

In the enterprise domain, Arakapis et al. \cite{Arapakis:p4l_p2p_telefonica:2023} introduced P4L, a private peer-to-peer learning system. As an asynchronous collaborative learning scheme, P4L allows users to participate in the learning process without depending on a centralized infrastructure. It ensures the confidentiality and utility of shared gradients employing strong cryptographic primitives. Also, it maintains resilience to user dropout and fault tolerance, highlighting the practical applicability and effectiveness of decentralized learning solutions in real-world settings. Finally, on the frontier of sixth-generation (6G) networks, Ridhawi et al. \cite{Ridhawi:digitaltwin_6g_framework:2023} proposed a decentralized zero-trust framework for digital twins. By integrating the zero-trust architecture into digital twin-enabled networks with DFL, they ensured the security, privacy, and authenticity of physical and digital devices. Their approach addresses the challenges of cooperation between devices and network components in a 6G environment, demonstrating the pivotal role of DFL in next-generation networks.

\section{Communications Threat Model in DFL}\label{sec:threatmodel}

The threat model primarily focuses on the communication aspects of DFL, presuming the co-existence of trusted participants who abide by network protocols and malicious participants who pose multilayered threats. The threat landscape in the communication channels of a DFL environment is complex, with malicious entities potentially playing passive or active roles. Passive malicious entities might eavesdrop on network communications, surreptitiously gaining access to sensitive information such as model parameters, aggregated gradients, or participants' metadata. In contrast, active malicious entities could actively interfere with network operations, manipulate data, introduce false information, or disrupt communication channels. These threats can originate from internal and external sources, with internal threats emerging from compromised or malicious network participants and external threats from entities outside the DFL topology.

As \change{outlined}{detailed} in Table~\ref{table:1}, a malicious participant \addtxt{in a DFL environment} can extract a wide range of sensitive information, each \change{with its distinct}{bearing unique implications} implications. \change{One example is model parameters, including the weights and biases for each neural network layer, encoding the knowledge the model has acquired. It's worth noting that while Homomorphic Encryption or Differential Privacy methods may prevent or obfuscate this extraction to some extent, the threat remains similar to the one faced in FL vanilla. The illicit acquisition of these parameters could enable a malicious actor to reconstruct the learning model, resulting in substantial data privacy breaches and potentially revealing sensitive insights. Furthermore, the topology offers valuable insights into the overall structure and interaction within the network. It can provide an adversary with knowledge of the structure, facilitating further targeted attacks.}{A notable example is the extraction of model parameters, such as weights and biases from each neural network layer, which encapsulate the learned knowledge of the model. Although methods like Homomorphic Encryption or Differential Privacy may impede or obscure this extraction, the underlying threat parallels that in FL. Unauthorized access to these parameters could allow a malicious entity to reconstruct the learning model, leading to significant data privacy violations and potentially exposing critical insights. Additionally, the network's topology provides valuable information about its structure and interactions, offering adversaries insights that could facilitate more targeted attacks.}

Additionally, the \change{roles assigned to participants within the DFL network could grant an adversary a comprehensive understanding of functional distribution and control mechanisms}{assigned roles within a DFL network can provide an adversary with a detailed understanding of the functional distribution and control mechanisms}. Unlike in FL vanilla, where all clients primarily hold the same role, this aspect of DFL architecture can aid an attacker in identifying which nodes to target for maximum disruption. Moreover, performance metrics and resource usage data could expose system vulnerabilities regarding performance and resource allocation strategies. An attacker might infer these metrics from the patterns and volume of network communications \cite{MartinezBeltran:fedstellar:2024}. Information about participant activity periods and the underlying model architecture could prove invaluable for an attacker. By analyzing communication timings and frequencies, an attacker might discern when specific nodes are most active, providing insights into the operational rhythms of the network. \change{Knowledge of the model architecture}{A deep understanding of the model architecture}, obtained through careful observation of network interactions and data exchanges, might expose the structure and operational logic of the model, thereby revealing potential weaknesses for exploitation. Finally, understanding communication patterns could prove beneficial for a malicious entity. \change{By scrutinizing the frequency and nature of participant interactions, an attacker could detect valuable patterns, forecast behaviors, and potentially impersonate trusted nodes, thereby gaining unauthorized access or sowing discord within the network.}{By examining the frequency and nature of participant interactions, an attacker could identify critical patterns, anticipate behaviors, and potentially impersonate trusted nodes to gain unauthorized access or disrupt the network.}

\begin{table*}[htb!]
\centering
\small
\caption{Information accessible to a malicious participant in DFL}
\begin{tabular}{p{3.7cm}p{11.6cm}}
\hline
\textbf{Information} & \textbf{Description} \\ [0.5ex] 
\hline
Model Parameters & Each layer $l_i$ in a model $M$ with $n$ layers has weight $w_i \in \mathbb{R}^{d_i \times d_{i-1}}$ and bias $b_i \in \mathbb{R}^{d_i}$, where $d_i$ is the number of neurons in layer $i$. The parameters of $M$ are the collection $\{w_i, b_i\}_{i=1}^{n}$. \\ 
\hline
Topology & The graph of the network $G(V,E)$, where $V$ is the set of vertices (participants) and $E$ is the set of edges (connections). If $V = \{v_1, v_2, ..., v_n\}$ and $E = \{(v_i, v_j) | v_i, v_j \in V, i \neq j\}$, the topology is fully connected.\\
\hline
Roles & Each participant $p_i \in V$ has a role $r_i \in \{$idle, trainer, aggregator, proxy$\}$. This can be mathematically represented by a function $R: V \rightarrow \{$idle, trainer, aggregator, proxy$\}$, where $R(p_i) = r_i$. \\
\hline
Metrics & Performance of the model (e.g., accuracy, precision, recall, F1 score) and resource usage (CPU, RAM, network) of the nodes. For resources, let $R$ be the resource, $U_R$ the usage, and $C_R$ the capacity. The usage rate is $R_{rate} = \frac{U_R}{C_R}$.\\
\hline
Activity Periods & If $T = \{t_1, t_2, ..., t_n\}$ represent the set of all time intervals and $A = \{a_1, a_2, ..., a_k\} \subseteq T$ the active intervals, then the activity ratio is $A_{ratio} = \frac{\sum_{i=1}^{k} a_i}{\sum_{i=1}^{n} t_i}$.\\
\hline
Model Architecture & A feedforward neural network with $n$ layers can be represented as a sequence of function compositions $f(x) = f_n(f_{n-1}(...f_2(f_1(x))))$, where $f_i(x) = \sigma(w_i \cdot x + b_i)$ is the operation for layer $i$, and $\sigma$ is the activation function.\\
\hline
Communication Patterns & If $M = \{m_{ij}\}$ is the set of all messages sent from participant $i$ to participant $j$, the frequency of communication between these participants can be quantified as $F_{ij} = \frac{|m_{ij}|}{\sum_{i,j}|m_{ij}|}$, where $|m_{ij}|$ is the number of messages exchanged. \\
\hline
\end{tabular}
\label{table:1}
\end{table*}

Numerous potential security threats can compromise the confidentiality, integrity, and availability of federated data and models. These threats primarily arise from the inherent vulnerabilities presented by the decentralization of learning processes and model sharing without the control of a central authority. The following communications threats have been identified (see \tablename~\ref{table:2}):

\newcommand{\noimportant}{{\color{ao}$\mathord{!}$}}
\newcommand{\important}{{\textbf{\color{amber}$\mathord{!!}$}}}
\newcommand{\critical}{{\color{red}$\mathord{!!!}$}}

\begin{table*}[htb!]
\caption{Attacks, goals, and information at risk in DFL}
\label{table:2}
\centering
\small
\begin{threeparttable}
\begin{tabular}{p{2.8cm}p{8cm}p{4cm}}
\hline
\textbf{Attack} & \textbf{Goal} & \textbf{Information at Risk} \\ 
\hline
Eavesdropping & Extract sensitive information to undermine integrity and security of the federated participants [\important] & $\sbullet[0.75]$ Model Parameters \newline $\sbullet[0.75]$ Topology \newline $\sbullet[0.75]$ Roles \\
\hline
MitM & Manipulate information or insert malicious data to disrupt federation operations [\critical] & $\sbullet[0.75]$ Communication Patterns \newline $\sbullet[0.75]$ Roles \\
\hline
Network Mapping & Know the network structure to launch more targeted future attacks on the federation [\noimportant] & $\sbullet[0.75]$ Topology \newline $\sbullet[0.75]$ Model Architecture \\
\hline
Eclipse Attacks & Isolate a node or group of nodes to extract information or disrupt DFL communications [\critical] & $\sbullet[0.75]$ Activity Periods \newline $\sbullet[0.75]$ Topology \newline $\sbullet[0.75]$ Roles \newline $\sbullet[0.75]$ Communication Patterns \\
\hline
\end{tabular}
\begin{tablenotes}
\item \noimportant\space Low importance,\space\important\space High importance,\space\critical\space Critical
\end{tablenotes}
\end{threeparttable}
\end{table*}

\begin{itemize}
    \item \textit{TH1. Eavesdropping}. In a DFL setting, an adversary could covertly monitor network communications or infiltrate a participant node to gain unauthorized access to sensitive data. This data could include model parameters, network topology, and participant roles. The adversary could then leverage this information to disrupt the federated process or impersonate a legitimate participant. This threat often persists undetected due to its covert nature, leading to prolonged periods of sensitive data leakage.

    \item \textit{TH2. MitM}. It involves an attacker intercepting and potentially manipulating the communication between two participant nodes. This enables the attacker to alter exchanged model parameters, introduce spurious data, or eavesdrop on the exchanged information, posing significant challenges to the integrity of the federated process.

    \item \textit{TH3. Network Mapping}. It aims to understand the structure of the federated network and the roles of participant nodes. By gaining this knowledge, attackers can predict and interfere with network operations, facilitating more targeted and potentially detrimental exploits.

    \item \textit{TH4. Eclipse}. This attack in DFL seeks to isolate a specific node or a group of nodes from the rest of the network. This isolation distorts the affected nodes' perception of the network state, causing them to act based on inaccurate information and potentially paving the way for additional security breaches.

\end{itemize}

In light of the identified threats, a comprehensive security module for DFL must account for these potential attack vectors and implement countermeasures to ensure robust operation and resilience against attacks. Crucially, achieving this goal involves striking a careful balance between enhancing security and managing the additional computational and network overhead that security measures may introduce.

\section{Security Module}\label{sec:module}

This section details the proposed security module, particularly examining its integration within a novel DFL platform and how it fortifies the network against a broad spectrum of cyber threats.

\subsection{Overview}

The security module comprises a set of cybersecurity strategies designed to safeguard the complex exchange of data and models in DFL. The distinctive features of DFL, such as decentralized aggregation, asynchronous communication, limited visibility to near neighbors, and participant independence, necessitate nuanced and versatile security measures. The limited visibility of DFL nodes, usually only to immediate neighbors, restricts the broader network anomaly detection. Participant independence complicates maintaining a secure environment as nodes decide when to commence model training or aggregation. This proposal responds to the growing need for advanced security mechanisms within the field of DFL, considering the diversity and sensitivity of data involved in these systems. This module employs sophisticated encryption methods and MTD techniques, making it highly adaptable to various DFL platforms:

\begin{itemize}
    \item \textit{Encryption}. Using a combination of symmetric and asymmetric encryption, the module ensures secure model exchanges and efficient key management. This strategy guarantees data confidentiality and provides robust protection against potential breaches.
    \item \textit{MTD Techniques}. These techniques, which include Neighbor Selection and IP/port switching, create a dynamic and unpredictable defensive layer within the system. By continuously changing communication pathways and nodes, these techniques make it increasingly difficult for potential attackers to gain a foothold in the system.
\end{itemize}

\subsection{Security Components}

The components of the security module comprise encryption techniques and MTD strategies. The encryption techniques, designed to ensure data confidentiality during the model exchange, combine the efficiency of symmetric encryption for data protection with the secure key management of asymmetric encryption. MTD techniques, such as Neighbor Selection and IP/port switching strategies, add a dynamic and shifting defensive layer to the system. These techniques introduce unpredictability and fluidity by continuously altering network communication pathways, making the system difficult for potential attackers to decipher due to the increased complexity and resource requirements for successful attacks. The integration of these components in a federated participant cycle within a DFL environment is depicted in Algorithm~\ref{alg:cycle}. This algorithm combines the elements of encryption and MTD, effectively creating a robust security layer within the DFL infrastructure.

\begin{algorithm}[ht!]
\caption{Federated participant cycle with Encryption and MTD Techniques in DFL}
\label{alg:cycle}
\footnotesize
\begin{algorithmic}[1]
\Require{$R$: local round, $\alpha$: learning rate, $\lambda$: regularization parameter, $S_j$: socket to neighbor j, $D$: local dataset, $E_{\text{sym}}$ / $E_{\text{asym}}$: symmetric/asymmetric encryption function, $D_{\text{sym}}$ / $D_{\text{asym}}$: symmetric/asymmetric decryption function, $MTD_{\text{IP}}$: IP/port MTD function, $MTD_{\text{N}}$: neighbor selection MTD function}

\Procedure{$MTD_{\text{N}}$}{$N_{\text{all}}, n$}
\Comment{\textbf{Neighbor Selection}}
    \State Initialize an empty list $N$
    \While{$|N| < n$}
        \State Select a neighbor $i$ from $N_{\text{all}}$ uniformly at random
        \If{$i \notin N$} 
            \State Add $i$ to $N$
        \EndIf
    \EndWhile
    \State \Return{$N$}
\EndProcedure

\Procedure{$MTD_{\text{IP}}$}{$config$}
\Comment{\textbf{IP/Port Switch}}
    \State Fetch a list of available IP and ports: $IP_{\text{avail}}, P_{\text{avail}}$
    \State Select a new IP and port from $IP_{\text{avail}}, P_{\text{avail}}$ uniformly at random
    \State Update $config$ with the new IP address and port
    \State \Return{$config$}
\EndProcedure

\State $D_{\text{Train}}, D_{\text{Test}} \gets split(D)$

\For{$r$ in $R$}
    
    \State $\theta \leftarrow Initialize()$ \Comment{\textbf{Initialize Local Model}}
    \For{each $(x, y)$ in $D_{\text{Train}}$}
        \State $\theta \leftarrow \theta - \alpha (\nabla_\theta J(\theta, x, y) + \lambda \theta)$ \Comment{\textbf{Train}}
    \EndFor
    
    \State $N \leftarrow MTD_{\text{N}}(N_{\text{all}})$
    
    \For{$j$ in $N$} \Comment{\textbf{Send}}
        \State $\theta_{\text{enc}} \leftarrow E_{\text{sym}}(\theta, K_{\text{sym}})$
        \State $K_{\text{sym\_enc}} \leftarrow E_{\text{asym}}(K_{\text{sym}}, K_{j_{\text{pub}}})$
        \State $\text{Send } \theta_{\text{enc}}, K_{\text{sym\_enc}} \text{ to } j \text{ via } S_j$ 
    \EndFor 
    
    \While{$not \; Timeout$}
        \For{$j$ in $N$} \Comment{\textbf{Receive}}
            \State $RP_{j_{\text{enc}}}, K_{j_{\text{sym\_enc}}} \leftarrow \text{Receive from } j \text{ via } S_j$ 
            \State $K_{j_{\text{sym}}} \leftarrow D_{\text{asym}}(K_{j_{\text{sym\_enc}}}, K_{\text{priv}})$
            \State $RP_j \leftarrow D_{\text{sym}}(RP_{j_{\text{enc}}}, K_{j_{\text{sym}}})$
        \EndFor
    \EndWhile
    
    \State $\theta \leftarrow \frac{1}{|N|+1} (\theta + \sum_{j \in N} RP_j)$ \Comment{\textbf{Aggregate} (FedAvg)}
    \State $\text{Update Local Model with } \theta$

\EndFor

\For{each $(x, y)$ in $D_{\text{\textbf{Test}}}$}
    \State $y_{pred} \leftarrow Predict(\theta, x)$ \Comment{\textbf{Test}}
    \State $L \leftarrow \frac{1}{|D_{\text{Test}}|}\sum_{i=1}^{|D_{\text{Test}}|} l(y_i, y_{pred_i})$ \Comment{\textbf{Compute Loss}}
\EndFor

\State $\text{Send metrics to controller}$ \Comment{\textbf{Report Metrics}}

\State $MTD_{\text{IP}}(config) \rightarrow config$

\end{algorithmic}
\end{algorithm}

\subsubsection{Communications Encryption}
\label{sec:encryption}

The integrity and confidentiality of the information exchanged among participants during the federation is a fundamental requirement in secure DFL systems. This security is achieved by combining symmetric and asymmetric encryption techniques, forming a comprehensive, multi-layered security infrastructure.

The first layer of this security architecture employs symmetric encryption. This method is computationally efficient and uses a single key for data encryption and decryption. The Advanced Encryption Standard (AES) algorithm, provided by the \textit{pycryptodome} library, is utilized for this layer. Known for its robust security and broad acceptance, the AES algorithm is an ideal choice, especially considering the resource constraints often present in many devices.

The second layer of the security architecture employs asymmetric encryption. This technique provides an additional layer of security by using a pair of keys: a public key for encryption and a private key for decryption. The RSA algorithm, also provided by the \textit{pycryptodome} library, is used for this layer. RSA eliminates risks associated with key sharing in symmetric encryption and ensures a secure channel for key exchange, protecting the symmetric keys used in the AES algorithm. \change{Key renewal, or the periodic updating of encryption keys, is another integral security feature of the module. The risk of a key compromise is significantly reduced by continuously renewing the keys during the federated process.}{Key distribution and management are central to this interconnected system, facilitated by the controller, which acts as a secure Key Distribution Center (KDC). Upon deployment, each node is authenticated by the controller (see Section~\ref{sec:fedstellar}) and issued digital certificates. This process underpins the trust and integrity of the public keys disseminated within the network. Moreover, the controller dynamically manages public key updates, scheduling regular key renewals in line with security protocols to swiftly address potential vulnerabilities.}

\subsubsection{MTD Techniques}

The MTD techniques \change{disrupt the attack surface of the system, increasing the difficulty for an attacker to exploit vulnerabilities and gain unauthorized access}{serve to obfuscate and alter the attack surface dynamically, posing a significant challenge for attackers attempting to exploit system vulnerabilities}. The proposed security module incorporates two MTD techniques: Neighbor Selection and IP/port switching.

The Neighbor Selection MTD technique minimizes network topology exposure to potential attackers. This technique can protect the nodes from targeted attacks by dynamically altering their communication partners in each learning cycle. By continually shifting the communication patterns in the network, the likelihood of an attacker successfully predicting or manipulating these patterns is significantly reduced. The random selection of neighbors is implemented using Python's built-in random library, ensuring unbiased and unpredictable selections for each cycle. The process for the Neighbor Selection MTD is fairly straightforward. In each federated round, a node randomly selects a subset of neighbors from all available participants (see Algorithm~\ref{alg:cycle}). This selection scheme is implemented using the socket library of Python, which provides low-level networking capabilities suitable for various network protocols, including TCP/IP, common in wired and wireless communications. The socket-based communication scheme offers reliability and flexibility, which are vital in a dynamic DFL environment.

The second technique is IP/port switching MTD, adding another layer of security. This method involves routinely changing the IP addresses and ports used by the federated nodes, further complicating the predictability of the attack surface. An attacker finds it difficult to sustain a prolonged attack on a specific node. In the proposed security module, IP/port switching is implemented by regularly selecting a new IP address and port from a pool of available ones. This selection is automated and randomized using the built-in capabilities of Python for network configuration. By dynamically altering the IP addresses and ports, the technique disrupts potential attackers' ability to predict the communication structure or execute targeted attacks.

\addtxt{Both techniques need to ensure uninterrupted and secure communication amid IP and port changes employing a rendezvous mechanism. To achieve this, the system implements a predictive notification mechanism. Before a node switches its network configuration, it broadcasts its neighbors an encrypted notification containing the new connection details. This notification, encrypted with the network's standard encryption protocols, allows each recipient node to update its records before the change. This decentralized approach eliminates the need for a real-time directory service and instead relies on the timely dissemination of IP/port updates directly between nodes. As a result, even when an IP address or port changes, the communicating nodes can independently reconcile the new configurations, thereby maintaining uninterrupted and secure connections. This method adheres to the principles of a decentralized network and reinforces the security infrastructure, ensuring the network resolution process remains robust against potential vulnerabilities.}

Building on the elaboration of the implemented security techniques, it is essential to understand their effectiveness, as depicted in \tablename~\ref{table:attacks-mitigations}. Encryption protects against eavesdropping, MitM, and eclipse attacks by protecting data during transmission. As a complement, MTD offers robust defenses against attacks such as Network Mapping or eclipse attacks.

\begin{table}[htb!]
\caption{Potential mitigations for attacks in DFL}
\label{table:attacks-mitigations}
\centering
\footnotesize
\begin{tabular}{cp{1.7cm}p{0.6cm}cp{0.8cm}}
\hline
\multicolumn{1}{c}{\makecell{\textbf{Security}\\\textbf{Components}}} & \multicolumn{4}{c}{\textbf{Attacks}} \\ 
\hline
& Eavesdropping & MitM & \makecell{Network\\Mapping} & Eclipse \\ 
\hline
Encryption & \makecell{\yestick} & \makecell{\yestick} & \makecell{\notick} & \makecell{\yestick} \\
\hline
MTD & \makecell{\notick} & \makecell{\yestick} & \makecell{\yestick} & \makecell{\yestick} \\
\hline
\end{tabular}
\end{table}

\subsection{Fedstellar Platform}
\label{sec:fedstellar}

Fedstellar is an innovative platform that facilitates the training of FL models across a wide array of physical and virtual devices \cite{MartinezBeltran:fedstellar:2024}. The platform is a hub for developing, deploying, and managing federated applications and provides a standardized approach for executing these processes. The architecture of Fedstellar is composed of three fundamental elements:

\begin{itemize}
    \item \textit{Frontend}. A user-centric interface that offers easy experiment setup and real-time monitoring, thus ensuring an intuitive user experience.
    \item \textit{Controller}. A central command unit orchestrates operations across the platform, ensuring seamless inter-module communication and efficient task execution.
    \item \textit{Core}. This critical component, deployed on each participating device, is responsible for vital functions such as model training and communication.
\end{itemize}

These components establish a robust and resilient architecture that provides sophisticated tools and metrics for federation management. This enables high transparency and efficiency in monitoring the learning process. Moreover, the platform contains extensible modules offering data storage, asynchronous capabilities, and effective model training and communication mechanisms.

\begin{figure*}[htb]
\includegraphics[width=0.9\textwidth]{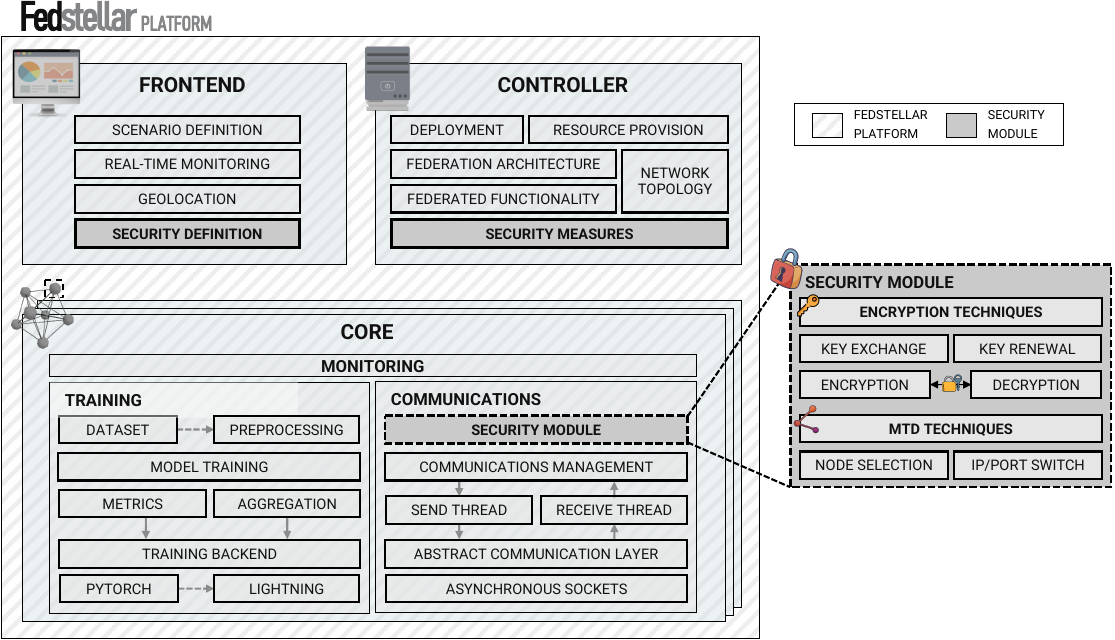}
\centering
\caption{Overall architecture of Fedstellar and the security module}
\label{fig:module}
\end{figure*}

The security module is integrated into the Fedstellar platform to demonstrate the proposed effectiveness and compatibility of the module. As depicted in Figure~\ref{fig:module}, the \textit{security module} is a pivotal functionality of the core component responsible for managing secure communications across the platform. Its integration into the core ensures robust protection for the vast and complex communication exchanges characteristic of DFL. To support the overall security structure, enhancements have also been made to the frontend and the controller components of the Fedstellar platform. 

\change{The}{In this sense, the} frontend encompasses the \textit{security definition} feature, enabling users to set and manage their security parameters conveniently. \change{Also}{Conversely}, the controller implements \textit{security measures}, a provision that efficiently manages and enforces the established security settings in real time. \addtxt{Also, it incorporates a participant authentication process based on JSON Web Tokens (JWT) during network deployment, conducted under encrypted communication (see Section~\ref{sec:encryption}). Upon joining the network, each node requests a token from the controller by providing its credentials. The controller validates these credentials and issues a JWT, which the node then uses for all subsequent communications within the network. This token, containing encrypted identity and permission information, ensures that only authenticated nodes participate in the network, enhancing security and preventing unauthorized access. The tokens have a limited lifespan, requiring nodes to periodically re-authenticate, thus maintaining ongoing network integrity.}

The integration of the security module maintains compatibility through its design, which leverages threaded processing for non-blocking operations and event passing between modules for effective communication. These provisions ensure that the addition of the module does not disrupt the existing functionalities of the platform but rather harmonizes with them, augmenting the capability of Fedstellar to efficiently manage diverse federations comprising various devices, network topologies, and algorithms.

\section{Validation Scenario}\label{sec:scenario}

The validation scenario of this study emulates an edge computing setting, which evaluates the performance of the proposed security module in a DFL environment. \change{As summarized in Table 4, the chosen configuration}{The validation was conducted in two distinct deployments: physical and virtual. The physical deployment, as detailed in \tablename~\ref{table:3},} encompasses a federation of eight physical devices: five Raspberry Pi 4 units and three Rock64 units. These devices are interconnected via a random network topology within the private local network. \addtxt{This topology is designed to mimic dynamic real-world environments, where connections between devices vary, offering insights into the federated process under fluctuating network conditions.} The Raspberry Pi 4 units, armed with a 1.5GHz quad-core 64-bit ARM Cortex-A72 CPU and 2GB of RAM, present a delicate balance between size, cost-effectiveness, and computational prowess, thereby rendering them a suitable choice for simulating edge nodes. The remaining three devices, Rock64 boards, enhance the system's heterogeneity by contributing slightly lower processing capabilities, characterized by a 64-bit ARM Cortex-A53 with a 1.5 GHz clock speed and up to 2GB RAM. \addtxt{To showcase the scalability of the solution, the experiment incorporates a virtual deployment comprising 50 Docker containers. Each container is configured to replicate the processing power and memory capacity of the physical devices. This expanded configuration offers a comprehensive testbed for evaluating the scalability and security module in a more complex DFL network. }\change{The deployed federation}{The physical and virtual deployment} \change{operates under the Fedstellar platform, and each participant utilizes the LeNet5 federated model trained on the MNIST dataset. The MNIST dataset was chosen for its relevance in many FL and pattern recognition research areas, making it a fitting choice for this validation scenario.}{is conducted on the Fedstellar platform, specifically designed to facilitate FL experiments. Within this platform, each participating node employs the LeNet5 neural network architecture. The choice of LeNet5 is strategic, given its relatively simple structure that allows for quick training and inference, thus suitable for DFL across devices with varying computational capabilities. The MNIST dataset is utilized to train and validate the federated models. Comprising 70,000 handwritten digits, MNIST provides a balanced and comprehensive dataset for benchmarking classification models.}

\begin{table}[htb!]
\caption{\change{Validation scenario using physical devices and eclipse attack}{Validation scenario using physical and virtual deployment}}
\label{table:3}
\centering
\footnotesize
\begin{tabularx}{\columnwidth}{XX}
\hline
\textbf{Characteristic} & \textbf{Description} \\ 
\hline
Participants & \circled{1} Physical deployment \newline $\sbullet[0.75]$ 5 Raspberry Pi 4 \newline $\sbullet[0.75]$ 3 Rock64 \newline \addtxt{\circled{2} Virtual deployment} \newline $\sbullet[0.75]$ \addtxt{50 Docker containers} \\
\hline
DFL Platform & Fedstellar \cite{MartinezBeltran:fedstellar:2024} \\
\hline
Federation Architecture & DFL \\
\hline
Network Topology & Random \\
\hline
Federated Model & LeNet5 \\
\hline
Dataset & MNIST \cite{Deng:MNIST:2012} \\
\hline
Security Configuration & \circled{1} Baseline \newline \circled{2} Encryption \newline \circled{3} Encryption and MTD \\
\hline
Attack &
Eclipse attack: \newline $\sbullet[0.75]$ One external attacker \newline $\sbullet[0.75]$ One target participant \\
\hline
\end{tabularx}
\end{table}

\begin{figure}[!htb]
\includegraphics[width=\columnwidth]{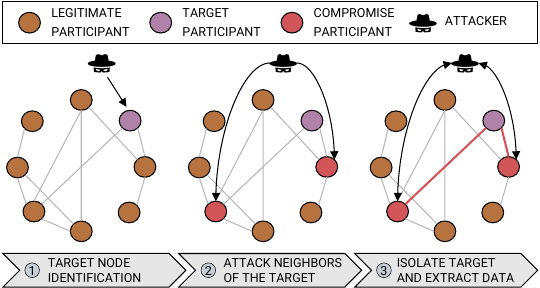}
\centering
\caption{\addtxt{Shematic representation of eclipse attack deployed in the validation}}
\label{fig:eclipse_attack}
\end{figure}

The security of the federation is assessed under three different configurations, providing an expansive view of its security posture under varied conditions. Initially, the federation functions with \circled{1} a baseline with no security measures and no malicious attack for subsequent security comparisons. Following this, the federation incorporates \circled{2} encryption techniques, forming its primary line of defense. Finally, the system operates with \circled{3} both encryption and MTD techniques, following the design of the proposed security module. To assess the resiliency of the \change{system}{security configuration} against cybersecurity threats, the validation scenario simulates an eclipse attack, a significant threat in decentralized networks \cite{Alangot:eclipse_attack_defense:2021, Niu:eclipse_attack:2022}. The choice of this attack is motivated by the number of security measures it requires, as shown in \tablename~\ref{table:attacks-mitigations}. The successful mitigation of this multifaceted attack in the validation scenario implies a high probability of successful defense against other potential attacks, as enumerated in \tablename~\ref{table:2}. \figurename~\ref{fig:eclipse_attack} shows the steps of the eclipse attack deployed: (i) involves isolating a chosen node, (ii) seizing control over its communications, and (iii) extracting valuable information. \change{For the simulation, two nodes are programmed to conduct the attack to quantify potential data theft risks and overall network vulnerability.}{The implementation of the eclipse attack, as detailed in Algorithm~\ref{alg:eclipse}, involved several technical considerations, particularly in network communication and manipulation. Initially, it required configuring two nodes to act as compromise participants. These nodes were set up using advanced socket programming techniques, allowing them to establish and hijack communication channels with the target node. By manipulating the routing tables and utilizing custom-built scripts, the attacking nodes were able to redirect traffic, effectively isolating the target node from the rest of the network. Following the steps outlined in \figurename~\ref{fig:eclipse_attack}, these nodes then took over the communication channels of the isolated node, using packet-sniffing tools and protocol spoofing to simulate data extraction processes.}

\begin{algorithm}
\caption{\addtxt{Implementation of eclipse attack in DFL}}
\label{alg:eclipse}
\footnotesize
\addtable{
\begin{algorithmic}[1]
\Require{$N$: Set of all nodes in the network, $T$: Target node, $A$: Attacker nodes}
\Ensure{Isolation and control over the target node $T$}

\State Initialize the network with nodes in $N$
\State Select target node $T$ from $N$
\State Initialize attackers in $A$
\For{each node $n$ in $N$} 
    \If{$n \in A$} \Comment{Node Identification}
        \State Begin monitoring communications of $T$
    \EndIf
\EndFor
\For{each communication link of $T$} \Comment{Node Isolation}
    \State Attacker nodes in $A$ intercept and block communications
\EndFor
\For{each outbound communication from $T$} 
    \State Redirect to attacker nodes in $A$ \Comment{Seizing Control}
\EndFor
\While{$T$ is isolated} \Comment{Information Extraction}
    \State Extract and analyze data from communications in $T$
    \State Attacker nodes mimic the legitimate network behavior
\EndWhile
\State \Return Success if $T$ remains isolated and controlled
\end{algorithmic}
}
\end{algorithm}

\section{Results}\label{sec:results}

\begin{figure*}[!htb]
  \centering
  \begin{subfigure}{.25\textwidth}
    \centering
    \includegraphics[width=\linewidth]{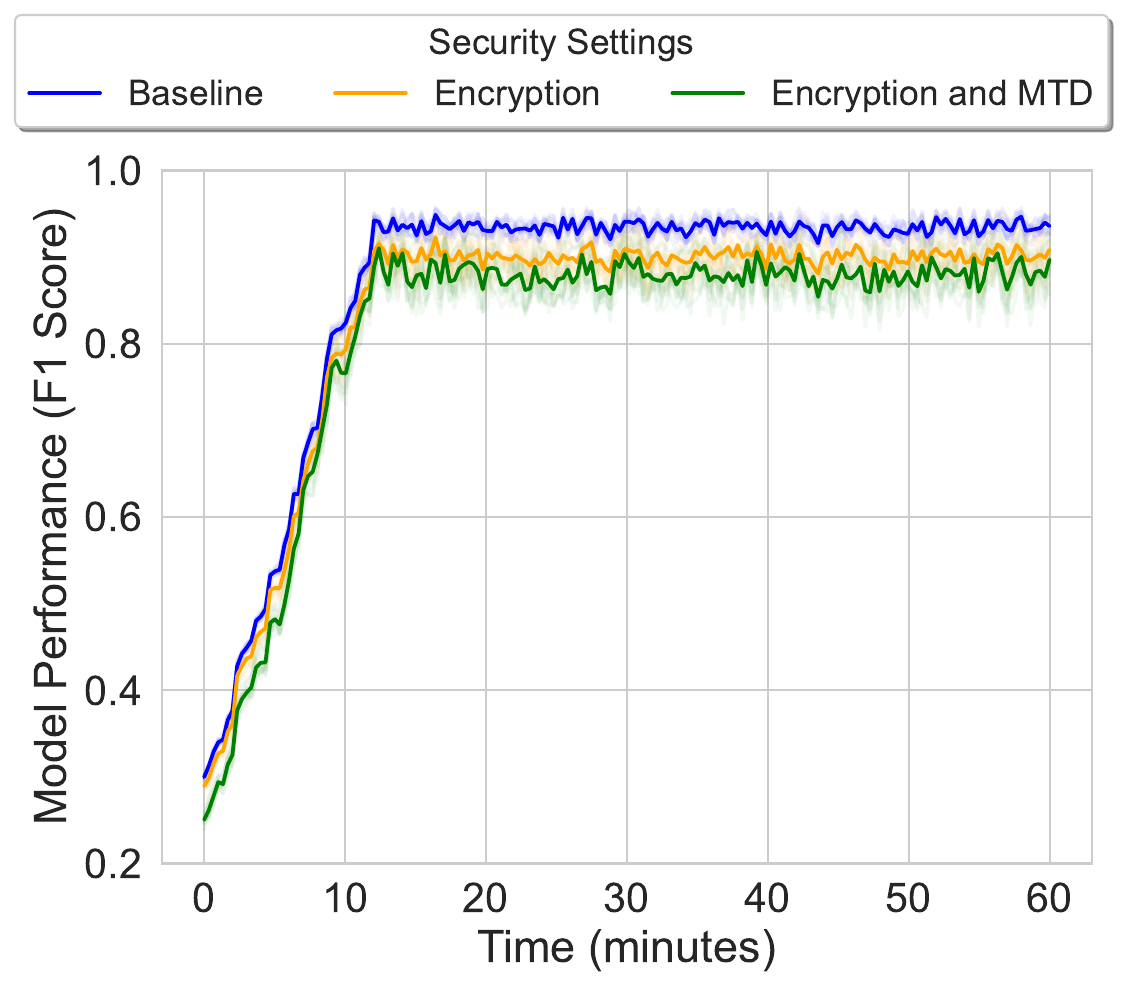}
    \caption{\addtxt{Model ($F_{1} \ score$)}}
    \label{fig:f1score-physical}
  \end{subfigure}%
  \begin{subfigure}{.25\textwidth}
    \centering
    \includegraphics[width=\linewidth]{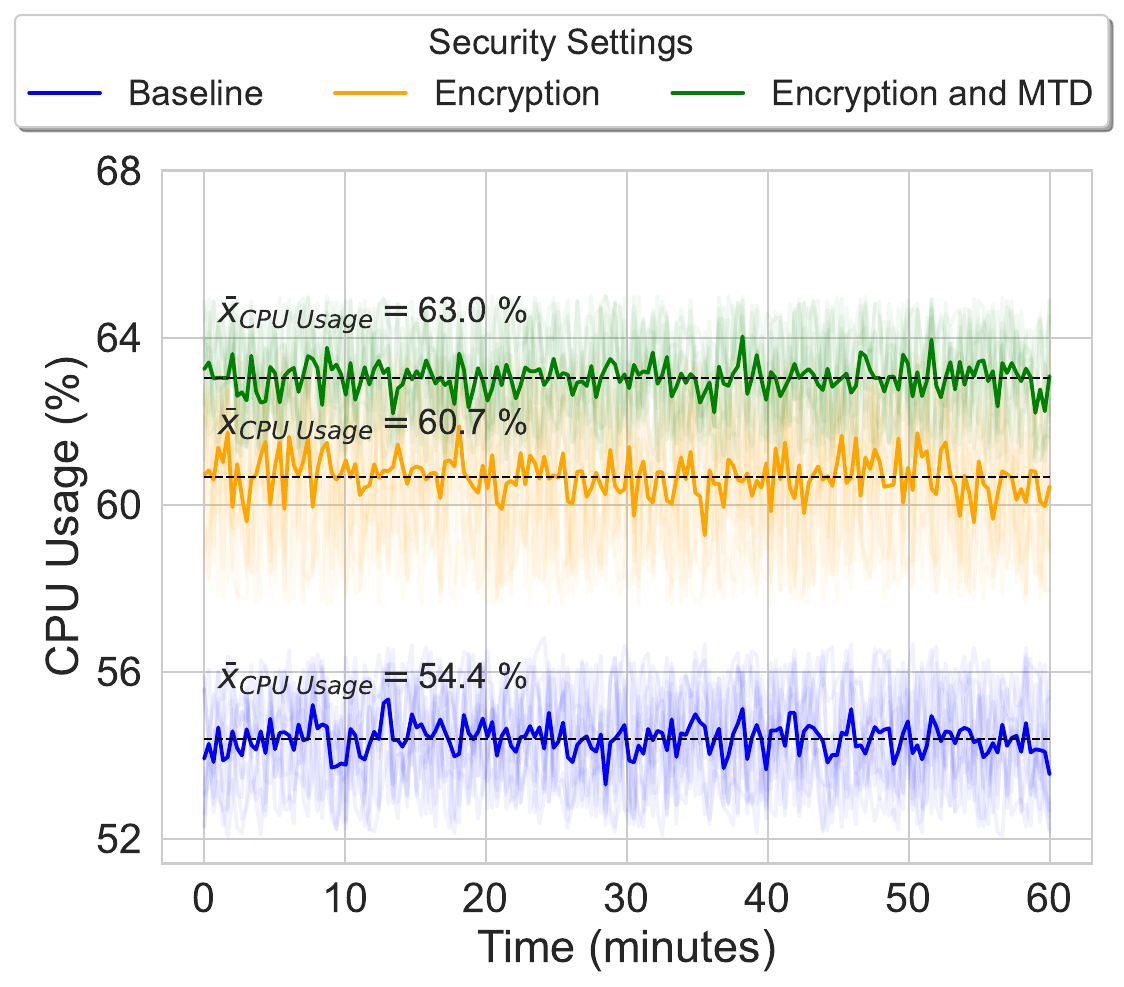}
    \caption{CPU usage $(\%)$}
    \label{fig:results_physical_cpu}
  \end{subfigure}%
  \begin{subfigure}{.25\textwidth}
    \centering
    \includegraphics[width=\linewidth]{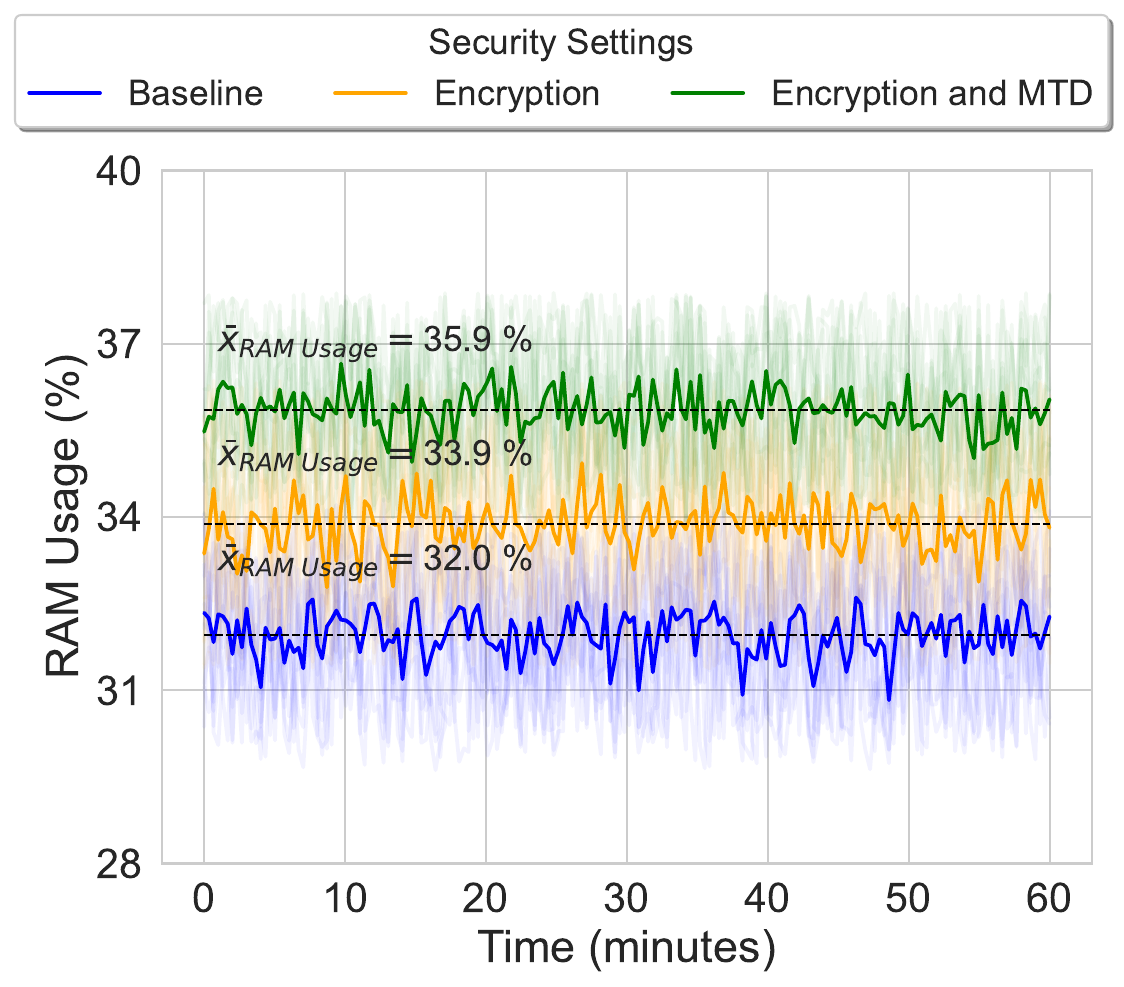}
    \caption{RAM usage $(\%)$}
    \label{fig:results_physical_ram}
  \end{subfigure}%
  \begin{subfigure}{.25\textwidth}
    \centering
    \includegraphics[width=\linewidth]{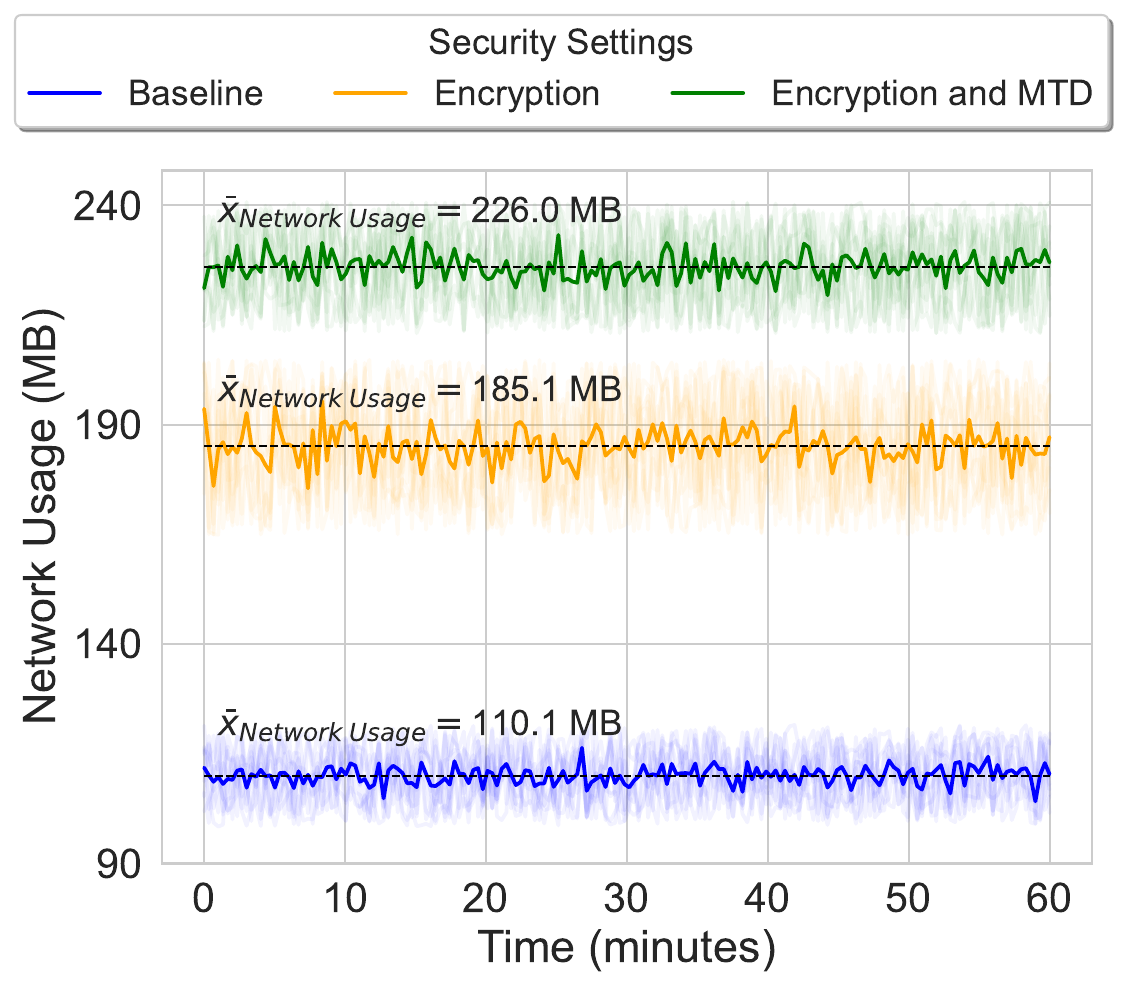}
    \caption{Network usage $(\text{MB})$}
    \label{fig:results_physical_network}
  \end{subfigure}

  \caption{Performance of Fedstellar in a physical deployment with eight participants using MNIST during 60 minutes}
  \label{fig:results_physical}
\end{figure*}

\begin{figure*}[!htb]
  \centering
  \begin{subfigure}{.25\textwidth}
    \centering
    \includegraphics[width=\linewidth]{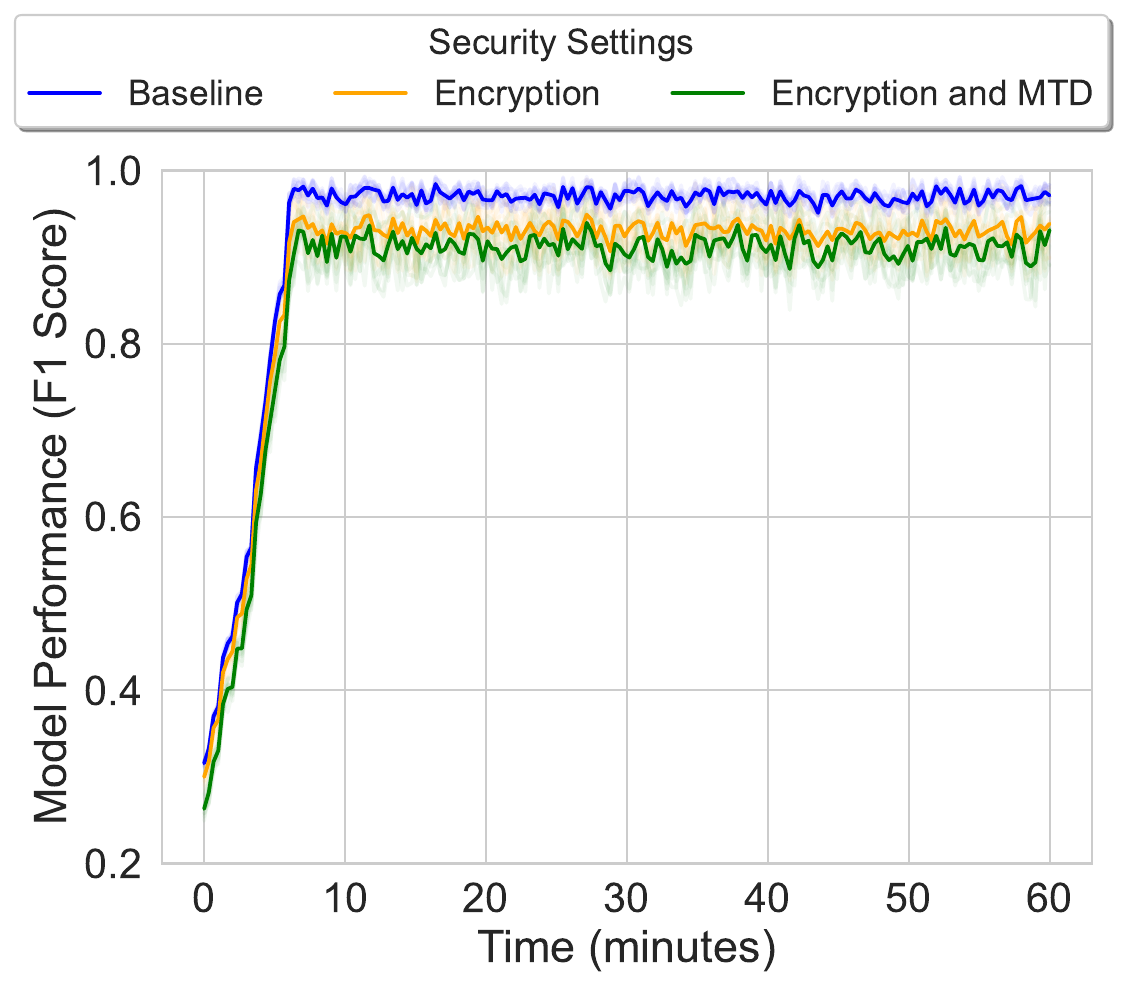}
    \caption{\addtxt{Model ($F_{1} \ score$)}}
    \label{fig:f1score-virtual}
  \end{subfigure}%
  \begin{subfigure}{.25\textwidth}
    \centering
    \includegraphics[width=\linewidth]{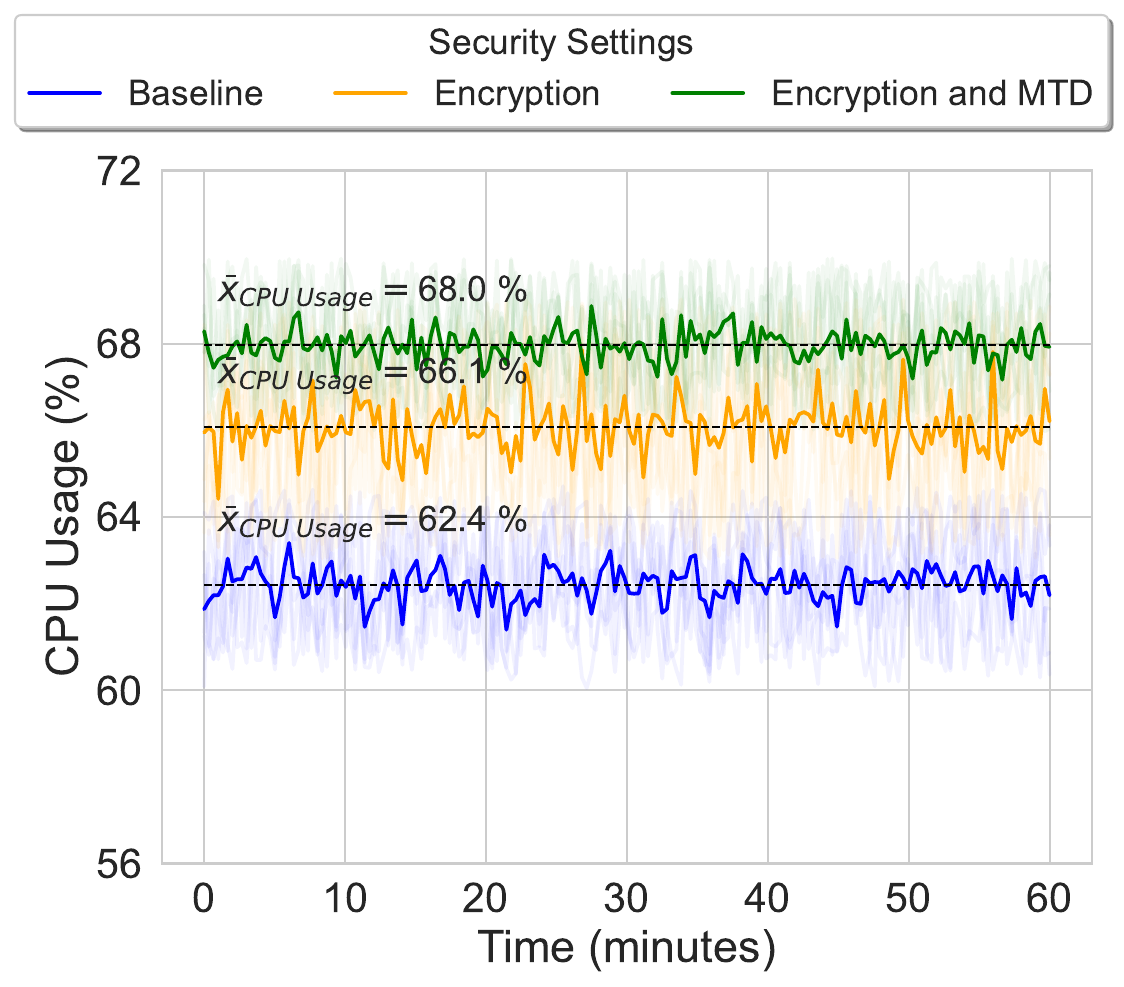}
    \caption{\addtxt{CPU usage $(\%)$}}
    \label{fig:results_virtual_cpu}
  \end{subfigure}%
  \begin{subfigure}{.25\textwidth}
    \centering
    \includegraphics[width=\linewidth]{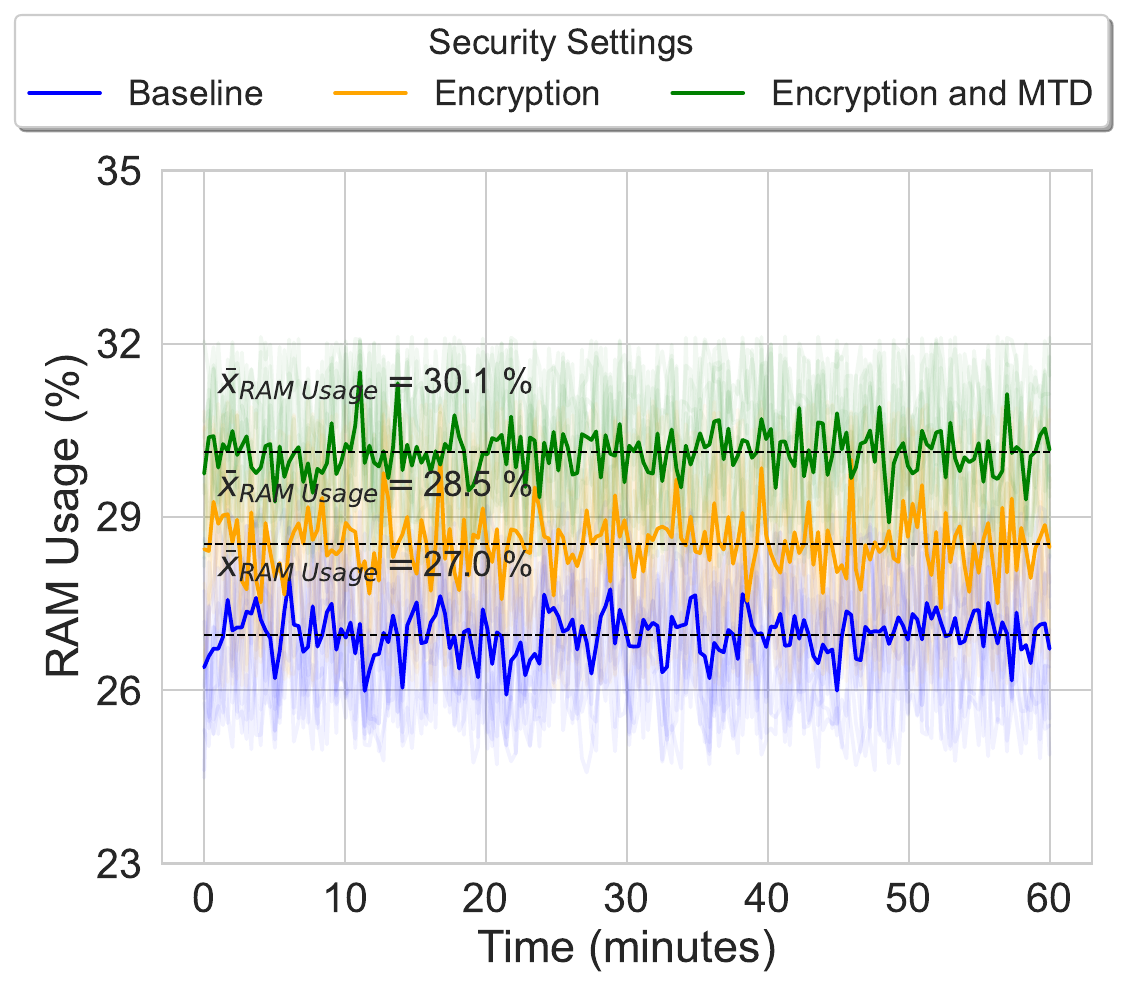}
    \caption{\addtxt{RAM usage $(\%)$}}
    \label{fig:results_virtual_ram}
  \end{subfigure}%
  \begin{subfigure}{.25\textwidth}
    \centering
    \includegraphics[width=\linewidth]{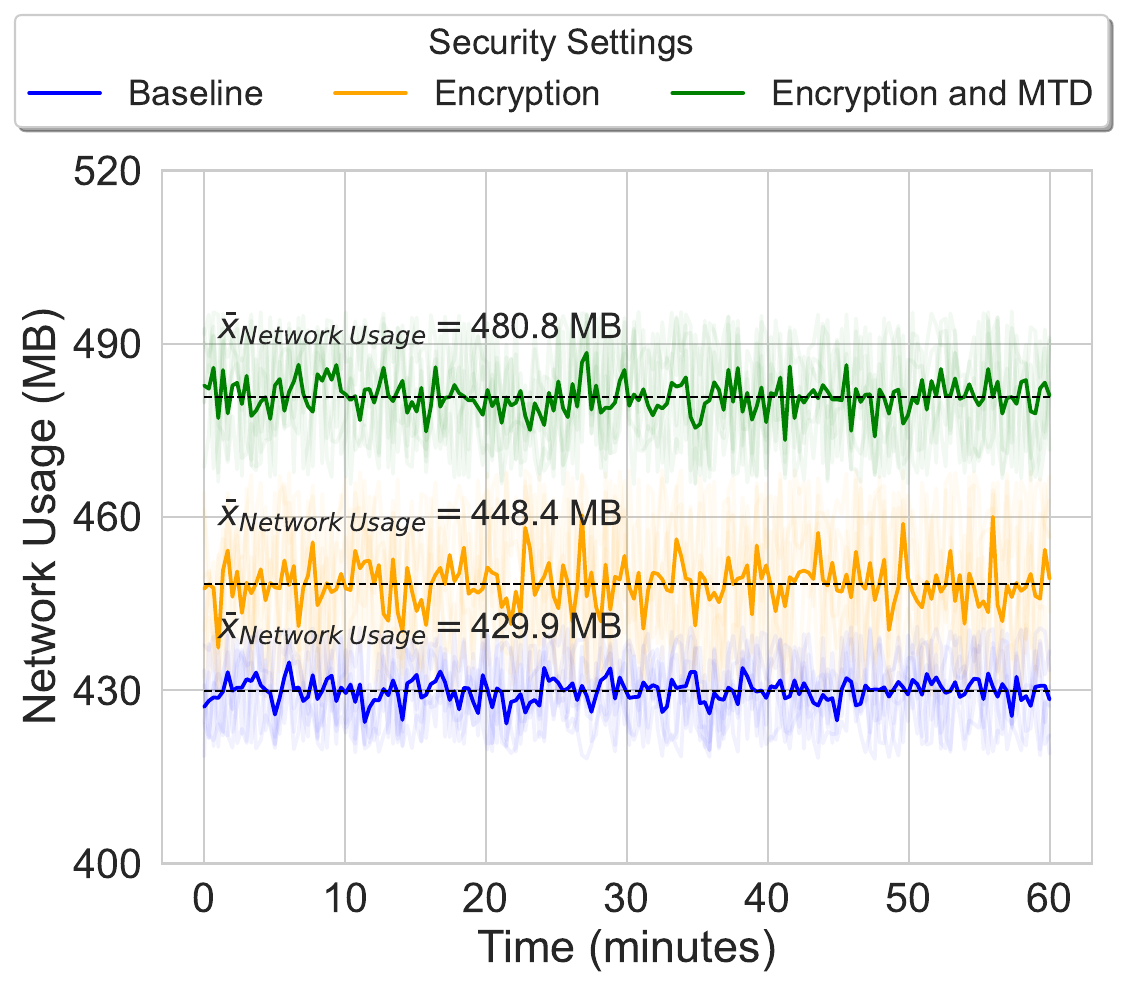}
    \caption{\addtxt{Network usage $(\text{MB})$}}
    \label{fig:results_virtual_net}
  \end{subfigure}

  \caption{\addtxt{Performance of Fedstellar in a virtual deployment with 50 participants using MNIST during 60 minutes}}
  \label{fig:results_virtual}
\end{figure*}

This section assesses the security module performance focused on performance indicators such as the $F_{1} \ score$ for federated models, the percentage of CPU and RAM usage, network traffic quantified in megabytes $(\text{MB})$, and model convergence time. \addtxt{\figurename~\ref{fig:results_physical} and \figurename~\ref{fig:results_virtual} show the performance indicators in the physical and virtual deployment, respectively.}

The diagram depicted in \figurename~\ref{fig:f1score-physical} demonstrates the average $F_{1} \ score$ for the federated models \addtxt{in a physical deployment}, under three separate security configurations: baseline without security techniques and malicious attacks on the network, encryption, and encryption combined with MTD techniques\addtxt{ to deal with attacks}. All three configurations exhibit a consistent growth pattern in the early stages of the federation process ($\approx$10 minutes). The baseline configuration continues upward, achieving an $F_{1} \ score$ of 97\%. This indicates the potential for high performance when security overheads are absent. However, when examining the configurations that include security measures, there is a slight decline in the $F_{1} \ score$. In the encryption configuration, the $F_{1} \ score$ peaks at 94\%, while in the combined encryption and MTD setting, the $F_{1} \ score$ fluctuates between 92.5\%. \addtxt{Similarly, \figurename~\ref{fig:f1score-virtual} shows the results in a virtual deployment. In this case, the growth is rapid in the first 6 minutes, with the baseline configuration reaching 98.9\%. The encryption configuration achieves an F1 score of 95.5\%, while the encryption with MTD configuration attains 93.8\%.} The variations observed throughout the federation process are likely due to the occasional computational overhead of the security mechanisms during the processing and transmission of data.

\change{A more granular view of the performance is provided by Fig. 3, which breaks down the performance into CPU, RAM, and Network metrics across the three security configurations. The observations corroborate the inherent trade-off between security and computational resources. The baseline configuration stands as the baseline, its usage determined by the computational demands of the training process, which averages at 54.6\% ($\pm$1.8\%). Upon the introduction of encryption, CPU usage is expected to rise due to the computational overhead associated with encrypting and decrypting communication. This is manifested in the encryption configuration where the CPU usage averages 60.9\% ($\pm$3.7\%). Further elevations in CPU usage are seen when encryption is combined with MTD. The additional processing required for managing dynamic communication routes leads to a higher average CPU usage of 63.2\% ($\pm$3.5\%).}{A more granular view of the performance in terms of CPU usage is provided by \figurename~\ref{fig:results_physical_cpu} and \figurename~\ref{fig:results_virtual_cpu} for physical and virtual deployments, respectively. In the physical deployment, the baseline CPU usage is 54.6\% on average, reflecting the computational load of the training process. With the introduction of encryption, there is an increase in CPU usage to 60.9\% due to the additional tasks of encrypting and decrypting data. These requirements further escalate when encryption is combined with MTD, leading to an average CPU usage of 63.2\%, attributed to managing dynamic communication routes. In contrast, the virtual deployment exhibits a different pattern, as shown in \figurename~\ref{fig:results_virtual_cpu}. The baseline configuration uses about 62.4\% of the CPU, which increases to 66.1\% with encryption, reflecting the computational overhead in a virtualized environment. Incorporating both encryption and MTD causes a CPU usage rise up to 68\%. These figures highlight the increased resource demands in the virtual deployment, particularly when the number of participants increased.}

\change{A similar pattern is observed for RAM usage.}{For RAM usage in both physical and virtual deployments, a discernible trend is evident, as highlighted in \figurename~\ref{fig:results_physical_ram} and \figurename~\ref{fig:results_virtual_ram}.} In the physical deployment, the baseline configuration exhibits a lower average of 31.9\%, reflecting the lower computational footprint when security measures are absent. However, including encryption mechanisms results in a slight increase in RAM usage due to the additional memory demands of the encryption process. Specifically, the encryption configuration averages 33.8\%, and when the MTD technique is added alongside encryption, the average RAM usage augments to 35.9\%. \addtxt{In contrast, the virtual deployment shows a different usage pattern. The baseline configuration in the virtual environment uses 27\% of the RAM, which is lower than in the physical deployment. This increases to 29.5\% with encryption and further to 31\% when both encryption and MTD are implemented.} This is attributable to the additional memory required for managing dynamic communication routes under MTD. Despite the marginal increase, it underscores the added resource requirements induced by security features.

\change{Furthermore, network traffic provides critical insights into the performance impacts of different security configurations. The average network traffic for the baseline}{Furthermore, network traffic, as depicted in \figurename~\ref{fig:results_physical_net} for the physical deployment and \figurename~\ref{fig:results_virtual_net} for the virtual deployment, provides critical insights into the performance impacts of different security configurations. In the physical setup, the baseline} configuration remains modest, averaging around 110.2 MB. However, the integration of security mechanisms leads to an increase in network usage. The encryption configuration generates an average of 185.2 MB of network traffic, while the encryption with MTD configuration pushes the average even higher, reaching 226 MB. \addtxt{In contrast, the virtual deployment, which involves a larger number of participants, exhibits higher network usage across all configurations. The baseline configuration shows an average network traffic of 429.9 MB, which rises to 448.4 MB with the implementation of encryption and further to 480.8 MB when encryption is combined with MTD techniques.}

\addtxt{Moreover, the detailed network metrics, as outlined in Table \ref{tab:network_metrics}, further elucidate the impacts of these security configurations on network performance. These metrics include (i) throughput, measuring data transmission efficiency; (ii) latency, indicating the communication speed; (iii) packet loss, reflecting data transmission reliability; and (iv) control overhead, representing the network cost due to security management. In physical deployments, the results show a slight decrease in throughput from 92 Mbps in the baseline to 85 Mbps with encryption and MTD, coupled with a gradual increase in latency and packet loss. Conversely, in virtual deployments, the throughput remains consistent across security settings, although lower than in physical setups, indicating a potential bottleneck in virtual environments. Interestingly, latency remains lower in virtual deployments compared to physical ones, possibly due to optimized routing in virtualized networks. However, packet loss and control overhead show a marked increase with more complex security configurations, emphasizing the additional network strain introduced by these security measures.}

\begin{table*}[!htb]
\caption{\addtxt{Network metrics under different security settings in DFL}}
\label{tab:network_metrics}
\centering
\small
\addtable{
\begin{tabular}{@{}llcccc@{}}
\hline
\makecell{\textbf{Deployment}} & \makecell{\textbf{Security}\\\textbf{Setting}} & \makecell{\textbf{Throughput}\\\textbf{(Mbps)}} & \makecell{\textbf{Latency}\\\textbf{(ms)}} & \makecell{\textbf{Packet Loss}\\\textbf{(\%)}} & \makecell{\textbf{Control Overhead}\\\textbf{(\%)}} \\ \hline
\multirow{3}{*}{Physical} & Baseline & 92 & 61 & 0.2 & 3.5 \\
 & Encryption & 87 & 63 & 0.5 & 4.7 \\
 & Encryption and MTD & 85 & 64 & 1.1 & 5.9 \\ \hline
\multirow{3}{*}{Virtual} & Baseline & 85 & 52 & 0.6 & 4.3 \\
 & Encryption & 81 & 52 & 0.6 & 7.1 \\
 & Encryption and MTD & 81 & 53 & 1.3 & 7.8 \\ \hline
\end{tabular}
}
\end{table*}

\change{As evidenced by Table 6, these results underscore an inherent tension in securing DFL. While deploying security protocols such as encryption and MTD is indispensable for safeguarding various aspects of the federated learning process, these measures invariably come with additional computational and network overheads. These escalated resource requirements, although a trade-off, provide a safeguard against the pervasive risk of data breaches and cyberattacks. This study thus offers an empirical guide, presenting the performance implications of various security configurations in real-world DFL scenarios. It showcases the balance between achieving high predictive accuracy and maintaining stringent security standards.}{As illustrated by Table \ref{table:results}, securing DFL systems with encryption and MTD techniques introduces notable computational and network overheads, evident in physical and virtual deployments. While these security measures increase CPU, RAM, and network usage, with virtual deployments showing higher resource utilization due to a larger number of participants, they are essential for protecting against data breaches and cyberattacks. This study highlights the critical balance in DFL between ensuring high predictive accuracy and adhering to rigorous security protocols, offering a comprehensive view of the performance trade-offs inherent in implementing robust security configurations in real-world scenarios.}

\begin{table*}[!htb]
\caption{\addtxt{Security settings, information protection, and performance in DFL. \textit{PD:} Physical Deployment, \textit{VD:} Virtual Deployment}}
\label{table:results}
\centering
\small
\begin{threeparttable}
\begin{tabular}{p{1.5cm}p{3.8cm}p{2.2cm}p{2.15cm}p{2.15cm}p{2.45cm}}
\hline
\makecell{\textbf{Security}\\\textbf{Settings}} & \makecell{\textbf{Information}\\\textbf{Protected}} & \multicolumn{4}{c}{\makecell{\textbf{Performance Metrics *}}} \\ 
& & \makecell{\textbf{F1 Score (\%)}} & \makecell{\textbf{CPU (\%)}} & \makecell{\textbf{RAM (\%)}} & \makecell{\textbf{Network (MB)}} \\
\hline
Baseline & \multirow[t]{2}{*}{N/A} & 
\textit{PD:} 97 $\pm$0.02 & \textit{PD:} 54.4 $\pm$8 & \textit{PD:} 32 $\pm$2.3 & \textit{PD:} 110.1 $\pm$12 \\ & & \addtxt{\textit{VD:} 98.9 $\pm$0.01} & \addtxt{\textit{VD:} 62.4 $\pm$12} & \addtxt{\textit{VD:} 27 $\pm$1.2} & \addtxt{\textit{VD:} 429.9 $\pm$8} \\
\hline

Encryption & \multirow[t]{2}{*}{\begin{minipage}[t]{3.8cm} $\sbullet[0.75]$ Model Parameters \\ $\sbullet[0.75]$ Roles \\ $\sbullet[0.75]$ Communication Patterns \end{minipage}} &
\textit{PD:} 94 $\pm$0.9 & \textit{PD:} 60.7 $\pm$7 & \textit{PD:} 33.9 $\pm$2.41 & \textit{PD:} 185.1 $\pm$21 \\ & & \addtxt{\textit{VD:} 95.5 $\pm$0.8} & \addtxt{\textit{VD:} 66.1 $\pm$9} & \addtxt{\textit{VD:} 29.5 $\pm$1.8} & \addtxt{\textit{VD:} 448.4 $\pm$20} \\ \\
\hline

\multirow[t]{5}{*}{\begin{minipage}[t]{1.5cm}Encryption and MTD \end{minipage}} & \multirow[t]{2}{*}{\begin{minipage}[t]{3.8cm} $\sbullet[0.75]$ Model Parameters \\ $\sbullet[0.75]$ Roles \\ $\sbullet[0.75]$ Communication Patterns \\ $\sbullet[0.75]$ Topology \\ $\sbullet[0.75]$ Activity Periods \end{minipage}} &
\textit{PD:} 92.5 $\pm$1.1 & \textit{PD:} 63 $\pm$7 & \textit{PD:} 35.9 $\pm$1.5 & \textit{PD:} 226 $\pm$15 \\ & & \addtxt{\textit{VD:} 93.8 $\pm$0.7} & \addtxt{\textit{VD:} 68 $\pm$9} & \addtxt{\textit{VD:} 31 $\pm$1.7} & \addtxt{\textit{VD:} 480.8 $\pm$18} \\ \\ \\ \\
\hline
\end{tabular}
\begin{tablenotes}
\item \addtxt{* Average values for each participant.}
\end{tablenotes}
\end{threeparttable}
\end{table*}

\section{Conclusion}\label{sec:conclusion}

This work formulated a threat model for DFL communications, providing a detailed understanding of potential security vulnerabilities and sensitive information that could be exposed during interactions between participating nodes. In response to these challenges, an innovative security module was developed for DFL communications. It incorporates robust defensive mechanisms, including symmetric and asymmetric encryption methods and MTD techniques, tailored to the unique structure and requirements of DFL. This security module was deployed within a real-world DFL framework called Fedstellar to evaluate its efficacy and practicality. \change{The validation scenario involved a random topology of eight physical devices solving an ML task using MNIST}{The validation scenario was conducted through two distinct deployments. The first involved a random topology of eight physical devices engaged in solving an ML task using the MNIST dataset} and facing \addtxt{a custom implementation of }eclipse attacks. \addtxt{Complementing this, a second deployment was executed in a virtual environment with 50 participants, expanding the scope and scale of the validation to a more extensive network scenario.} \change{This scenario}{Both deployments} allowed the module to be rigorously evaluated under three security configurations: baseline \change{(no security)}{without security and malicious attacks}, encryption, and a composite of encryption and MTD. \change{The assessments validated the performance of the proposed module, demonstrating a satisfactory F1 score of 93\% on average, with an acceptable rise in system overhead. The peak values for CPU usage, network traffic, and RAM utilization were 63.2\% ($\pm$3.5\%), 230 MB ($\pm$15 MB), and 33.9\% ($\pm$1.5\%), respectively, thereby demonstrating the efficiency and practicality in real-world DFL applications.}{The assessments validated the performance of the proposed module across both physical and virtual deployments, demonstrating an average F1 score of approximately 93\% with an acceptable increase in system overhead. The peak values observed in the physical deployment for CPU usage, network traffic, and RAM usage were 63\% ($\pm$7\%), 226 MB ($\pm$15 MB), and 35.9\% ($\pm$1.5\%), respectively. In the virtual deployment, these metrics slightly increased due to the larger scale of operation, reaching 68\% ($\pm$9\%) for CPU usage, 480.8 MB ($\pm$18 MB) for network traffic, and 31\% ($\pm$1.7\%) for RAM usage. These results demonstrate the efficiency and practicality of the security module in diverse DFL applications, accommodating various deployment scales and complexities.}

Future research could consider developing and integrating new security techniques into the current security module to enhance the resilience of DFL environments further. Researchers might assess these enhancements across dynamic network topologies and more participant devices to better understand their efficacy in real-world, large-scale applications. Additionally, simulations with a wider variety of potential attacks would provide valuable insights into the robustness of these defensive methods under diverse threat scenarios. These advancements could significantly contribute to achieving secure, efficient, and scalable deployment of DFL.

\backmatter

\section*{Declarations}

\begin{itemize}
\item Competing interests: This work has been partially supported by \textit{(a)} 21629/FPI/21, Fundación Séneca, Región de Murcia (Spain), \textit{(b)} the strategic project DEFENDER from the Spanish National Institute of Cybersecurity (INCIBE) and by the Recovery, Transformation and Resilience Plan, Next Generation EU, \textit{(c)} the Swiss Federal Office for Defense Procurement (armasuisse) with the DEFENDIS and CyberForce projects (CYD-C-2020003), and \textit{(d)} the University of Zürich UZH.
\item Availability of data and materials: Data sharing does not apply to this article as no datasets were generated during the current study.
\end{itemize}

\bibliography{sn-bibliography}

\end{document}